\documentclass[runningheads]{cl2emult}

\usepackage{makeidx}  
\usepackage{graphicx} 
\usepackage{subeqnar} 
\usepackage{multicol} 
\usepackage{cropmark} 
\usepackage{lnp}      
\makeindex            

\begin{document}
\title*{NIR spectroscopy with the VLT of a sample of ISO selected 
Hubble Deep Field South Galaxies}
\toctitle{NIR spectroscopy with the VLT of a sample
\protect\newline of ISO selected Hubble Deep Field South Galaxies}
%
%
\titlerunning{VLT Spectroscopy of ISOHDFS galaxies}
%
\author{Dimitra Rigopoulou \inst{1}
\and Alberto Franceschini\inst{2}
\and Reinhard Genzel\inst{1}
\and Paul van der Werf\inst{3}
\and Herv\'e Aussel\inst{2}
\and Catherine Cesarsky\inst{4}
\and Michel Dennefeld\inst{5}
\and Seb Oliver\inst{6}
\and Michael Rowan-Robinson\inst{6}}
\authorrunning{Dimitra Rigopoulou et al.}
%
%
\institute{Max-Planck-Institut f\"ur extraterrestrische Physik, 
 Postfach 1603, 85740 Garching, Germany
\and Dipartimento di Astronomia, Viccolo Osservatorio 5 I-35122,
 Padova, Italy
\and Leiden Observatory, PO Box 9513, 2300 RA, Leiden, The Netherlands
\and European Southern Observatory, Karl-Schwarzschild-str. 2, 85740 Garching, 
 Germany
\and Institut d'Astrophysique de Paris - CNRS,98bis Boulevard Arago, 
 75014 Paris, France
\and Imperial College of Science Technology and Medicine, 
 Astrophysics Group, Blackett Laboratory, Prince Consort Rd., 
 London, SW2 1BZ, U.K.}

\maketitle              

\begin{abstract}
A new population of faint galaxies characterized by an extremely high
rate of  evolution with redshift up to z$\sim$1.5 has recently been
discovered by ISO. These sources are likely to contribute
significantly to the cosmic far-IR extragalactic background.  We have
carried out near-infrared VLT-ISAAC spectroscopy  of a sample of
ISOCAM galaxies  from the Hubble Deep Field South. The rest-frame
R-band spectral properties of the ISO population resembles that of
powerful dust-enshrouded active starburst galaxies.

\end{abstract}

\section{Introduction}

IRAS observations show that
about 30\% of the total starlight is emerging in the mid and
far-infrared (Soifer and Neugebauer 1991). About 25\% of the 
high mass star formation within 10 Mpc originates in dusty infrared 
luminous galaxies.
IRAS counts indicate evidence for evolution at low flux levels in
(Ultra)luminous Infrared Galaxies ((U)LIRGs, L$_{bol} >$ 10$^{12}$ 
L$_{\odot}$, Lonsdale et al. 1990, Franceschini et al. 1988).
However, the IRAS survey only sampled the local Universe (z$<$0.3) and the 
extrapolation of this evidence for cosmic evolution to higher 
redshifts is speculative. Since then, almost
all information about high-z galaxies relied on optical surveys.
Although these surveys have been succesful in discovering distant galaxies 
(e.g. Steidel et
al. 1996) and constraining the star formation history of the Universe
(Madau et al. 1996) they were able to tell only part of the story. 
The COBE detection
of an extragalactic infrared background (Puget et al. 1996) with an
integrated intensity similar or higher than that of  optical light
(Hauser et al. 1998, Lagache et al. 1999) suggests that a  significant
part of the star formation in the Universe is obscured and thus missed
by the various optical surveys.

The advent of the Infrared Space Observatory (ISO) has had a very
significant
impact on the studies of high -z star forming galaxies.  Operating in
the sensitive to dust and PAH emission mid-infrared regime,  ISOCAM on
board ISO, was more than 1000 times more sensitive than IRAS and  thus
had the potential to study infrared bright galaxies at redshifts beyond 0.5.
A number of cosmological surveys have been performed using ISOCAM ranging
from  large and shallow ones to pencil-beam deep ones reaching down to
a few $\mu$Jy sensitivities (see Cesarsky et al. this volume). The
source counts from all  ISOCAM surveys combined with those of IRAS are
in good agreement with a no evolution model ($\alpha$= -2.5) up to a
flux level of 100 mJy. However at fainter flux levels the situation
rapidly changes: the counts lie an order of magnitude higher than the
predictions of 
no-evolution models. This steepening in the log N--log S plot at $\sim$ 0.4 mJy 
implies
that ISOCAM surveys have probably revealed a new population of
strongly evolving galaxies (Elbaz et al. 1999)

Among the deepest surveys performed by ISOCAM are the observations of
the Hubble Deep Field North (N) and South (S) regions (Williams et al.
1996) resulting in the detection of $\sim$ 150 sources down to $\mu$Jy
levels.  The next major step is to explore the nature of the ISOCAM
population with optical$/$near-IR spectroscopy.  The results we
discuss in this work focus on a sample of galaxies drawn from the
ISOCAM survey of the HDF-S field. We present the results of
near-infrared  spectroscopic followup carried out using the Very Large
Telescope (VLT) in an attempt to characterize the nature of this new
population.

\section{ISOCAM Observations of HDF-S}

The Hubble Deep Field S was observed by ISOCAM as  part of the
European Large Area ISO Survey (ELAIS, see Oliver et al, this
volume). The observations were carried out at two wavelengths, LW2(
6.75 $\mu$m) and LW3(15 $\mu$m). The data have been analysed
independently by Oliver et al. (2000) and Aussel et al. (2000). The
latter analysis was carried out using the PRETI
method (Starck et al. 1999)
and resulted in the  detection of 63 sources in the LW3
band. The results presented here are based on the Aussel et al. (2000)
analysis.

\section{The VLT ISO-Hubble Deep Field-S sample}

The sample presented here was selected from the ISOCAM LW3 detections
of the Hubble Deep Field-S (hereafter ISOHDFS sample). For the sample
selection we imposed two criteria: a) a secure LW3 detection and b) a
secure counterpart in the I band image (Dennefeld et al. 2000), or a
counterpart in the K band image (EIS Deep, DaCosta et al. 1998). 
The latter was
necessary because in a number of cases there were more than just one I
band counterparts for one ISOCAM candidate.  We did not apply any
selection based on colours. Our sample (hereafter VLT ISOHDFS sample)
is thus a fair representation of the strongly evolving ISOCAM
population near the peak of the differential source counts (Elbaz et
al. 2000).

Our VLT ISOHDFS sample contains about 25 galaxies. The LW3 flux ranges
between 100--400 $\mu$Jy.  For the near-infrared observations we used
spectroscopic  redshifts from optical spectroscopy where available for
z$<$0.7 (Dennefeld et al. 2000) or photometric redshifts  estimated
based on the model PEGASE (Fioc and Rocca-Volmerange 1997).   Our
photometric redshift determination turned out to be accurate to $\pm$0.1 and
provided  a very powerful tool for ISAAC followup.

\section {VLT Observations}

The observations were carried out during 1999 September 20-24 with the
infrared spectrometer ISAAC (Moorwood et al. 1998) on ESO telescope
UT1,  on Paranal in Chile. For the observations we used the low
resolution grism R$_{s} \sim $ 600 and a slit of 1''.  The choice of
filter was dictated by our aim to detect H$_{\alpha}$ line
emission. Based on our spectroscopic and$/$or photometric redshift
estimates we chose the equivalent Z, SZ, J or H filter. To maximise the
observing efficiency we positioned the slit (which had a 2$^{\prime}$
length) in such a way as to include at least two galaxies at any given
orientation.

We observed 13 galaxies and H$_{\alpha}$ was succesfully
detected in all but  two of them. [NII] emission is also seen in some
spectra.  Finally, in some cases we have also detected emission from
the [SII] $\lambda\lambda$ 6717 6731 lines.  The H magnitudes of the
observed galaxies varied in the range 19-22 mag.

Most of the spectra were acquired with a 1hr on source integration
time.  With this integration time we were able to detect emission
lines as well as continuum in almost all of the galaxies. Observations
of spectroscopic  standard stars were also performed to allow proper
flux callibration of the galaxy spectra.  The spectra were reduced in
part using the ECLIPSE software (Devillard 1998) and standrard IRAF
routines.

Spectra of two of the galaxies observed are shown in Figures 1 and 2.
\begin{figure}
\centering \includegraphics[width=.8\textwidth]{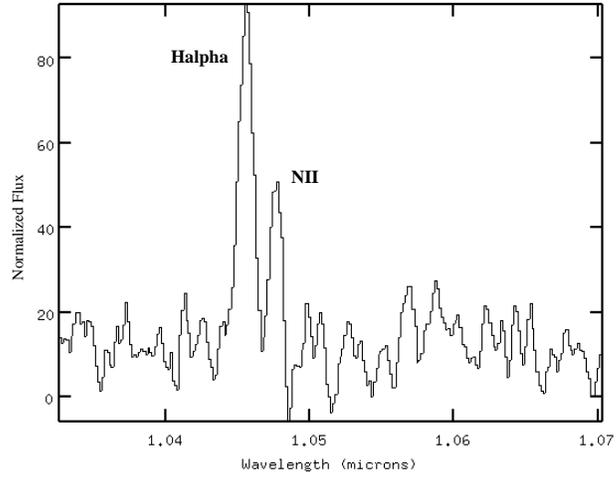}
\caption[]{Near-infrared spectrum of ISOHDFS 53, showing the clear
detection of the H$_{\alpha}$ and NII emission line. The spectroscopic
redshift of this object is 0.58}
\label{s53}
\end{figure}

\begin{figure}
\centering \includegraphics[width=.8\textwidth]{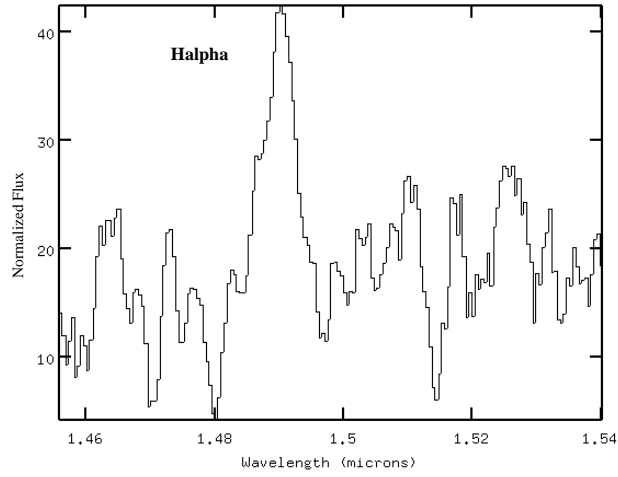}
\caption[]{ Near-infrared spectrum of ISOHDFS 39, showing the clear
detection of the H$_{\alpha}$ emission line. The spectroscopic
redshift of this object is 1.29}
\label{s39}
\end{figure}

\section{The nature of the ISOCAM faint galaxies}

Prior to our study no near-infrared (rest-frame R-band) spectroscopy
had been carried out for the ISOCAM population, primarily because of
the faintness of the galaxies. Optical spectroscopy (rest-frame
B-band) has been done for HDF-N (Aussel et al. 1999a) and the Canada
France Redshift  Survey (CFRS) field (Flores et al. 1999). The ISOCAM
HDF-N galaxies have been cross-correlated with the optical catalog of
Barger et al. (1999) resulting in 38 galaxies with confirmed
spectroscopic redshifts. Flores et al. have identified ~22 galaxies
with confirmed spectroscopic information. In both of these samples the
median redshift is about 0.7. Our VLT ISOHDFS sample contains 1 galaxy
with z$<$0.5, 6 galaxies 0.5$<$z$<$0.75  and 6 galaxies with
0.9$<$z$<$1.3.  Thus our sample has a z-distribution 
very similar to the HDF-N (Aussel et al.) and CFRS (Flores et al.) samples.

Rest-frame B-band spectra  host a number of emission and absorption lines 
related
to the properties of the  starburst in a galaxy. Based on these
features galaxies can be classified  according to their starburst
history.  Strong H$_{\delta}$ Balmer absorption and no emission lines
are characteristic of k$+$A galaxies.  The presence of significant
higher level Balmer absorption lines implies the presence of a
dominating A-star population. Such an A-star population may have been
created in a burst 0.1--1 Gyr ago (post-starbursts).  The simultaneous
presence of Balmer absorption and  moderate [OII] emission, known as
e(a) or E$+$A galaxies, may be characteristic of somewhat younger, but
still post-starburst systems or, alternatively, active but highly dust
absorbed starbursts.  As we will show, the ISOCAM galaxies are in fact
powerful starbursts hidden by large amounts of extinction.

Despite the fact that the local mid-infrared luminosity function is
dominated by AGNs (Fang et al. 1998), both Aussel et al. and Flores et
al. find  a large fraction of the ISOCAM galaxies are predominantly
powered by star formation.  Based on spectrophotometric SEDs of 19
galaxies Flores et al. deduce that the majority of the CFRS ISOCAM
population exhibits post-starburst characteristics with the star
formation occuring about 1 Gyr prior to the observed event.  

\subsection{ISOHDFS Galaxies are dusty luminous active starbursts}

As discussed in the previous section k$+$A galaxies can be naturally
explained as post-starbursts with a prominent burst 0.1--1.0 Gyr ago.
E(a) (or E$+$A) galaxies would then equally naturally be explained as being
somewhat younger systems. B-band spectra of local dusty starbursts
such as M82 (L$\sim$ 10$^{10}$ L$_{\odot}$, Kennicutt et al. 1992),
LIRGs (L$\sim$ 10$^{11}$ L$_{\odot}$, Wu et al. 1998), and ULIRGs
(L$\sim$ 10$^{12}$ L$_{\odot}$, Liu and Kennicutt 1995), surprisingly
enough, look like e(a), yet H$_{\alpha}$ is strong (typical for an HII
or liner galaxy). The apparent discrepancy between these two
signatures can be reconciled: differential dust extinction is at play.

Large amounts of dust exist within the HII regions where  the
H$_{\alpha}$ and [OII] line emission originates. [OII] emission is
affected more than H$_{\alpha}$ simply because of its shorter
wavelength.  The continuum is due to A-stars.  This A-star signature
comes from an earlier (0.1--1.0 Gyr) star formation epoch that is not
energetically dominant, in fact plays a small role once the dusty
active starburst is dereddened.  Such a scenario implies that these
galaxies undergo multiple burst events: the less extincted
population is due to an older burst while in the heavily dust
enshrouded HII regions there is ongoing star formation.

A notable property of the e(a) galaxies appears to be their low
EW(OII)$/$  EW(H$_{\alpha}+$NII) ratio. Such low ratios  have already
been observed in the spectra of distant clusters (Dressler et
al. 1999) and nearby mergers (Poggianti 99). Similar behaviour is
found among the dusty Luminous Infrared Galaxies (LIRGs) studied by Wu
et al. (1998) or the ULIRGs studied by Liu and Kennicutt (1995). The
behaviour of the (OII)$/$(H$_{\alpha}+$NII) ratio is shown in the
EW(OII)--EW(H$_{\alpha} +$ NII) diagram (Figure 3): The majority of
the points lie below the straight line, which represents a fit for
normal field galaxies at  low redshifts. Since for our ISOHDFS sample
there aren't any B-band spectra available yet we use the values
reported for the CFRS galaxies (median 20$\pm$15, Flores et al. 1999).
It follows from Figure 3 that the ISOHDFS galaxies occupy the same
region in the EQW(OII)$/$EQW(H$_{\alpha} +$ NII) diagram as the LIRGs
and ULIRGs.

\begin{figure}
\centering \includegraphics[width=1.\textwidth]{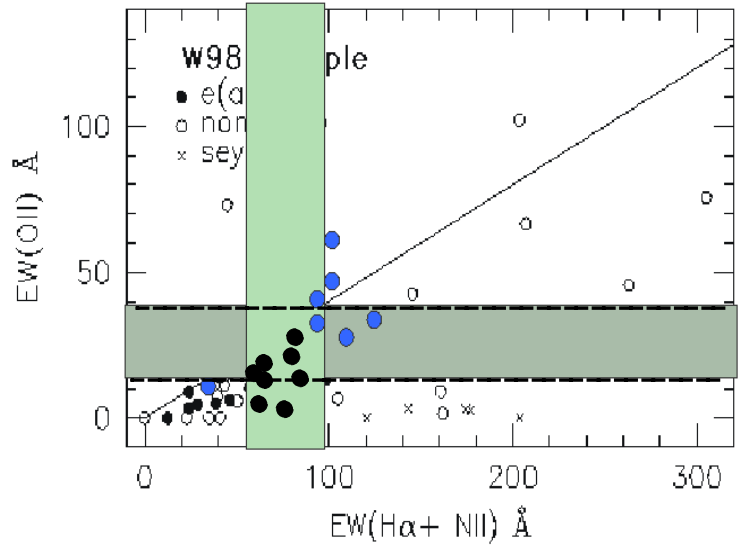}
\caption[]{EW(OII)--EW(H$_{\alpha} +$ NII) diagram (from Poggianti and
Wu (1999), filled circles e(a) galaxies, open circles non-e(a) galaxies,
crosses Seyferts, grey filled circles ULIRGs).
The shaded bars represent the area of the VLT ISOHDFS
galaxies. The intersection of the two bars is the location of our
galaxies, indistiguishable from the dusty Luminous e(a) galaxies.

}
\label{oiiha}
\end{figure}

\section{Star Formation Rates}

In ionization bounded HII regions the Balmer emission line
luminosities scale  directly with the ionizing fluxes of the embedded
stars. It is thus possible to use the Balmer emission lines to derive
quantitative star formation rates in galaxies (e.g. Kennicutt
1983). Since H$_{\alpha}$ is typically the strongest of the Balmer
lines is thus the best for such applications. However, beyond
redshifts of  z~0.2-0.3 H$_{\alpha}$ redshifts into the near-IR which
is not easily accessible with 4m class telescopes.  Due to this
constraint several workers (e.g. Dressler et al. 1995) have relied  on
measurements of the [OII] $\lambda$3727 doublet as a star formation
index for distant galaxies.

The conversion factor between ionizing flux and star formation rate
(SFR) is usually computed using an evolutionary synthesis model. Only
massive stars ($>$ 10 M$_{\odot}$) with short lifetimes (a few million
years) contribute significantly to the integrated ionizing flux.
Based on our recent measurements of the H$_{\alpha}$ line strength and
following the recipe of Kennicutt (1998):

SFR(M$_{\odot}/yr$) = 7.9 $\times$10$^{-42}$L(H$_{\alpha}$)(erg
s$^{-1}$).

Using this formula we estimate that the SFR rates in the
ISOHDFS galaxies range between 10-40 M$_{\odot}/$yr, corresponding to total
luminosities of $\sim$1--4$\times$10$^{11}$ L$_{\odot}$ (assuming H$_{o}$ = 75 
km/s/Mpc, $\Omega$=1).
These values imply that
the ISOCAM galaxies would fall in the LIRG class.
However, the present SFR
estimates can be used only as an indicative {\em lower limit} of the real
SFR in these galaxies. An upward correction of a factor of 3 should be
applied to correct for the extinction (assuming A$_{\small V} \sim$1.5).

A preliminary calculation of SFR rates based on the LW3 fluxes
(Franceschini et al. 2000) shows indeed that the SFR is 3-4 times 
higher than estimated from H$_{\alpha}$ measurements.

\section{Conclusions}

We have presented first results of a followup program aiming to
characterize the nature of the strongly evolving population discovered
by the ISOCAM surveys.  The detections of strong H$_{\alpha}$ emission
and large EW in almost all of the galaxies we observed implies that
these objects are active star-bursts. Our results imply that the star
forming regions inside these galaxies are affected by differential
extinction which is responsible for the low EW(OII)$/$EW(H$_{\alpha}$)
often observed in these galaxies.  This result demonstrates that it is
very dangerous to derive star formation rates from UV data alone since
these wavelengths are susceptible to higher extinction.  Thus a
significant fraction of star formation could may have been missed by optical
surveys.

Using the observed H$_{\alpha}$ emission lines we estimate that the
SFR rate in the VLT ISOHDFS galaxies ranges between 10 -40 M$_{\odot}
/$ yr. This SFR rate serves as a lower limit to the true SFR in these
galaxies.  The SFR as derived from ISOCAM observations is about 3
times higher. We conclude that the ISOCAM surveys have unveiled a new
population of active dusty starbursts which probabaly account for 
a substantial fraction of the FIR$/$submm background (see Elbaz et al. 2000).


\clearpage
\addcontentsline{toc}{section}{Index}
\flushbottom
\printindex

\end{document}